\documentstyle[aps,prl,multicol,psfig]{revtex}
\tighten
\begin{document}
\title{Influence of incommensurate dynamic charge-density wave
scattering on the line
shape of superconducting high-T$_c$ cuprates}
\author{G. Seibold$^{\dagger}$ and M. Grilli$^{*}$}  
\address{$^{\dagger}$ Institut f\"ur Physik, BTU Cottbus, PBox 101344, 
         03013 Cottbus, Germany}
\address{$^*$ Istituto Nazionale di Fisica della Materia e
Dipartimento di Fisica, Universit\`a di Roma ``La Sapienza'',\\
Piazzale A. Moro 2, 00185 Roma, Italy}

\maketitle

\begin{abstract}
We show that the spectral lineshape of superconducting 
La$_{2-x}$Sr$_x$CuO$_4$ (LSCO) and Bi$_2$Sr$_2$CaCu$_2$O$_{8+\delta}$
(Bi2212) can be well described by
the coupling of the charge carriers to collective incommensurate  
charge-density wave (CDW) excitations.
Our results imply that besides antiferromagnetic (AF) fluctuations also 
low-energy CDW modes can contribute to the observed
dip-hump structure in the Bi2212 
photoemission spectra. In case of underdoped LSCO 
we propose a possible interpretation of ARPES data
in terms of a grid pattern of 
fluctuating stripes where the charge and spin scattering
directions deviate by $\alpha=\pi/4$. Within this scenario we find that 
the spectral intensity along $(0,0) \rightarrow (\pi,\pi)$ is strongly
suppressed consistent with recent photoemission experiments.
In addition the incommensurate charge-density wave scattering 
leads to a significant
broadening of the quasiparticle-peak around $(\pi,0)$.
\end{abstract}

\vspace*{0.2cm}

\begin{multicols}{2}
\section{Introduction}
There is an increasing experimental evidence \cite{TALLON,BOEB}
that the peculiar properties
of the superconducting cuprates, both in the normal and the
superconducting phase are related to the occurrence 
of a Quantum Critical Point (QCP) located near (slightly above)
the optimal doping. These evidences corroborate the consensus
that quantum criticality related to some ``exotic'' \cite{EXOTICQCP}
or more traditional \cite{SACHDEV,PINES,CAST,PERALI,CAPECOD} type of ordering
is at work in these systems. Accordingly 
the phase diagram of the cuprates is naturally 
partitioned in a (nearly) ordered, a quantum critical, and 
a quantum disordered region naturally corresponding to
the under-, optimally, and over-doped regions of the phase
diagram of the cuprates respectively. 
In this framework a key open issue is to determine the type of
ordering which is established (at least in a local and slowly dynamical
sense) in the underdoped phase of the cuprates. While the direct
detection of the exotic order is more difficult and elusive, it is
quite natural that the spin and charge fluctuations associated
to the more traditional proposals \cite{SACHDEV,PINES,CAST,PERALI}
affect in a rather apparent way the physical properties. 
Especially one would expect that the modification of the electronic 
structure due to dynamical spin- or charge scattering should be observable
in angle-resolved photoemission spectroscopy (ARPES).
In fact, ARPES data of high-T$_c$ Bi$_2$Sr$_2$CaCu$_2$O$_{8+\delta}$ 
(Bi2212) compounds show several specific features in the superconducting
state which can be interpreted in terms of a coupling of the electrons
to a collective mode \cite{NORMAN2}.
Below $T_c$, these spectra are characterized by an unsual lineshape around
$(\pi,0$) (M-point) which consists of a sharp peak at low energy 
followed by a hump at
higher energies. Both features are separated by a dip.
In addition the sharp peak persists at low energy over a large
region in k-space whereas the hump correlates well with the underlying
normal state dispersion \cite{NORMAN1}.
Since the energy difference between dip and peak position is similar
to the energy of the $(\pi,\pi)$ resonance observed in inelastic 
neutron scattering \cite{NEUTRONS} is has been suggested 
that the mode responsible for the ARPES features corresponds
to collective spin fluctuations \cite{CAMPU}. 
Although this proposal is quite appealing one needs to understand
by which mechanism the substantial spin fluctuations can be sustained 
in optimally and overdoped materials. In fact, since doping
rapidly disrupts the antiferromagnetic (AF) order and with the
occurrence of an AF-QCP in the very underdoped region where
the Ne\'el temperature vanishes, spin scattering should be
a minor effect at large dopings. This difficulty can 
be overcome by considering the (traditional) ordering in the underdoped
phase to be driven by the charge. Then the occurrence of (a local
and possibly slowly dynamical) charge ordering is naturally
accompanied by substantial spin fluctuations since the spin coupling
resurrects in the hole-depleted regions. This strong spin and charge
coupling is also an effective  mechanism driving the weak charge
modulation around the second-order quantum transition to strongly
anharmonic charge (and spin) {\em stripes} in the deeply underdoped materials.
The relevant role of charge ordering has a solid experimental support
at least in some classes of cuprates.
First evidence for incommensurate stripe structures
in the high-T$_c$ compounds was given by neutron scattering 
experiments \cite{TRAN} in Nd-doped lanthanum cuprate. Here 
the Low-Temperature Tetragonal (LTT) lattice structure
is suitable to effectively pin the charge stripes giving
rise to static stripe ordering so that both magnetic and charge-order
peaks become detectable. However, also in Nd-free materials 
including the YBCO compounds where
the magnetic incommensurate peak has been observed \cite{YAM,ARAI} 
the simultaneous occurence of charge order has been claimed
in Ref. \cite{MOOK} and from
NQR measurements in Refs. \cite{TEITEL,BRINKMAN,MEHRING,HUNT}.
Moreover, it has has been argued \cite{TALLON} that the QCP can be
deduced from the incommensurate neutron peak intensity which
vansishes in the slightly overdoped regime  x$\approx 0.19$.
This coincides surprisingly well with experiments in
La$_{2-x}$Sr$_x$CuO$_4$ (LSCO) compounds where superconductivity has
been suppressed by pulsed magnetic fields \cite{BOEB} showing an underlying
metal-insulator transition at about the same critical doping.
Recent photoemission experiments \cite{INO,INO2,ZHOU} on superconducting 
underdoped La$_{2-x}$Sr$_x$CuO$_4$ (LSCO) have also revealed 
a broad quasiparticle peak (QP) around the $(\pi,0)$ point. On the other
hand no QP can be identified along the $(0,0) \rightarrow (\pi,\pi)$
direction in contrast to the Bi2212 compound where along the
diagonals a clear Fermi surface crossing has been observed.
Exact diagonalization studies of the t-t'-t''-J model with an 
additional phenomenological stripe potential \cite{TOHYAMA} 
have demonstrated that these features may be due to the
coupling of the holes to  vertical charged (static) stripes.

In the present paper we want to focus on the dynamic aspect of 
incommensurate charge-density wave (CDW) scattering within the
framework of a QCP scenario.
In this context we discuss the differences between the  
photoemission spectra of Bi2212 and LSCO respectively. 
Especially we show that the different features can be explained 
by the assumption that in Bi2212 the charge carriers
couple to an incommensurate CDW oriented along the $(1,0)-$ and $(0,1)-$
directions whereas in LSCO the dynamic CDW scattering is along the diagonals. 
The results presented below supplement considerations of Ref. \cite{GOETZ}
where we have shown that a two-dimensional 'eggbox-type' charge-pattern
can reproduce the essential features of the normal state Bi2212 Fermi surface,
namely the reduction of spectral weight around the M-points associated with 
the opening of a pseudogap in these regions of k-space.
Here we extend our considerations to the case of LSCO assuming
that the k-dependence of the electron self-energy is an essential 
ingredient in the description of ARPES data. 

We note that in a recent paper \cite{ESCHRIG} 
Eschrig and Norman have analyzed ARPES data within a model of electrons
interacting with a magnetic resonance using a similar approach than in the
present paper. Comparing their results with those reported below
one can see that AF and
CDW scattering have similar effects on the electronic states around the
M-points and the similarity between our results and those
in Ref. \cite{ESCHRIG} provides a support to the idea that
spin and charge fluctuations coexist and cooperate in determining the
spectral properties.

After having introduced the formalism in Sec. II we
will present in Sec. III the analysis of ARPES spectra for both
LSCO and Bi2212 materials. We finally conclude our discussion in Sec. IV.

\section{Formalism}
We consider a system of superconducting electrons exposed to
an effective action
\begin{equation}
S=-\lambda^2 \sum_q  \int_0^{\beta}d\tau_1\int_0^{\beta}d\tau_2
\chi_q(\tau_1-\tau_2) \rho_q(\tau_1) \rho_{-q}(\tau_2)
\end{equation}
describing dynamical incommensurate CDW fluctuations.
Using Nambu-Gorkov notation the unperturbed Matsubara Greens function matrix
$\underline{\underline{G}}$ is given by
\begin{eqnarray}
G^0_{11}(k,i\omega)&=& \frac{u_k^2}{i\omega-E_k}
+\frac{v_k^2}{i\omega+E_k}\\
G^0_{22}(k,i\omega)&=&\frac{v_k^2}{i\omega-E_k}+\frac{u_k^2}{i\omega+E_k}\\
G^0_{12}(k,i\omega)=G^0_{21}(k,i\omega)&=&-u_k v_k\left\lbrack 
\frac{1}{i\omega-E_k}-\frac{1}{i\omega+E_k}\right\rbrack
\end{eqnarray}
where the BCS coherence factors are defined as $u_k^2=\frac{1}{2}
(1+\frac{\epsilon_k-\mu}{E_k})$ and $v_k^2=\frac{1}{2}
(1+\frac{\epsilon_k+\mu}{E_k})$ respectively.
The leading order one-loop contribution to the self-energy reads
as
\begin{equation}\label{SELF}
\underline{\underline{\Sigma}}(k,i\omega)=-\frac{\lambda^2}{\beta}
\sum_{q,ip} \chi_q(ip)
\underline{\underline{\tau_z G}}^0(k-q,i\omega-ip)
\underline{\underline{\tau_z}}
\end{equation}
which in turn allows for the calculation of $\underline{\underline{G}}$
via
\begin{equation}
\underline{\underline{G}}=\underline{\underline{G}}^0+
\underline{\underline{G}}^0 \underline{\underline{\Sigma}}
\underline{\underline{G}}.
\end{equation}
It should be mentioned that the self-energy eq. (\ref{SELF}) differs 
from the analogous expression for the coupling to spin-fluctuations 
by the $\underline{\underline{\tau_z}}$ Pauli matrices.
Finally the spectral function can be extracted from 
$A_k(\omega)=Im G_{11}(k,\omega)$.

Note that our approach differs from the standard Elisahberg treatment
by the fact that already the unperturbed system
displays coherent superconducting order. Since the incommensurate 
CDW fluctuations have been shown to be strongly attractive in the
d-wave channel \cite{PERALI} the idea is to incorporate this feature 
already on the zeroth order level by a frequency independent d-wave 
order parameter. Thus the self-energy is gapped by 
$\underline{\underline{G}}^0$, however,
for our present considerations
this approximation is sufficient since we are interested in line shape
phenomena occuring at energies $\omega \gg \Delta^{SC}$. 

\section{Results}
In order to simplify the calculations we consider a Kampf-Schrieffer-type
model susceptibility \cite{KAMPF}
which is factorized into an $\omega$- and q-dependent part, i.e.
\begin{equation}
\chi_q(i\omega)=W(i\omega)J({\bf q})
\end{equation}
where $W(\omega)=-\int d\nu g(\nu) 2\nu/(\omega^2+\nu^2)$
is some distribution of dispersionless propagating bosons and
\begin{equation}
J({\bf q})=\frac{{\cal N}}{4}
\sum\limits_{\pm q_x^c;\pm q_y^c}\frac{\Gamma}{\Gamma^2
+2-\cos(q_x-q_x^c)-\cos(q_y-q_y^c)}.
\end{equation}
${\cal N}$ is a suitable normalization factor introduced to
keep the total scattering strength constant while varying $\Gamma$.
The above susceptibility 
contains the charge-charge correlations which are enhanced at the four
equivalent critical wave vectors $(\pm q^c_x,\pm q^c_y)$.
As we will show below the static limit $g(\nu)=\delta(\nu)$ 
together with an 
infinite charge-charge correlation length $\Gamma \rightarrow 0$
allows one to relate the coupling parameter $\lambda$ to the
parameter set of the Hubbard-Holstein model with long-range
Coulomb interaction by following the approach in Ref. \cite{GOETZ}.
Concerning the normal state dispersion we use a tight-binding fit
to normal state ARPES data \cite{NORMAN3,KIM}
including the hopping between first, second and third nearest neighbors,
i.e. $\epsilon_k = t(\cos(k_x)+\cos(k_y))/2 +t'\cos(k_x)\cos(k_y)
+t''(cos(2k_x)+cos(2k_y))/2$. For Bi2212 we take $t=-0.59, t'=0.164,
t''=-0.052$ and for LSCO $t=-0.35, t'=0.042,t''=-0.028$.
The BCS gap is assumed to be d-wave symmetry  $\Delta_{SC}({\bf k})
=\Delta_0(\cos(k_x)-\cos(k_y))$ where $\Delta_0$(Bi2212)$=32$meV and
$\Delta_0$(LSCO)$=10$meV respectively.

\subsection{Bi$_2$Sr$_2$CaCu$_2$O$_{8+\delta}$}
In Ref. \cite{GOETZ} we have already demonstrated that a static
two-dimensional eggbox-type charge modulation with ${\bf q_c}$
oriented along the vertical directions can account for the basic 
Fermi surface (FS) features in the optimally and underdoped Bi2212 compounds.
Here we supplement these considerations by a detailed analysis
of the photoemission lineshape in the superconducting state.  
Concerning the frequency dependent part of the susceptibility
we restrict ourselves to the simplest case of a single energy 
excitation, i.e. we choose $g(\nu)=\delta(\nu-\omega_0)$.

The coupling of electrons with a dispersionless bosonic mode
is a well known subject in solid state theory and has been
intensively investigated in \cite{ENGELSBERG}.
In the present context the new feature is a) the inclusion of
superconductivity and b) the role of a strongly q-dependent coupling
between electrons and the incommensurate CDW mode.

\begin{figure}
{\hspace{0.cm}{\psfig{figure=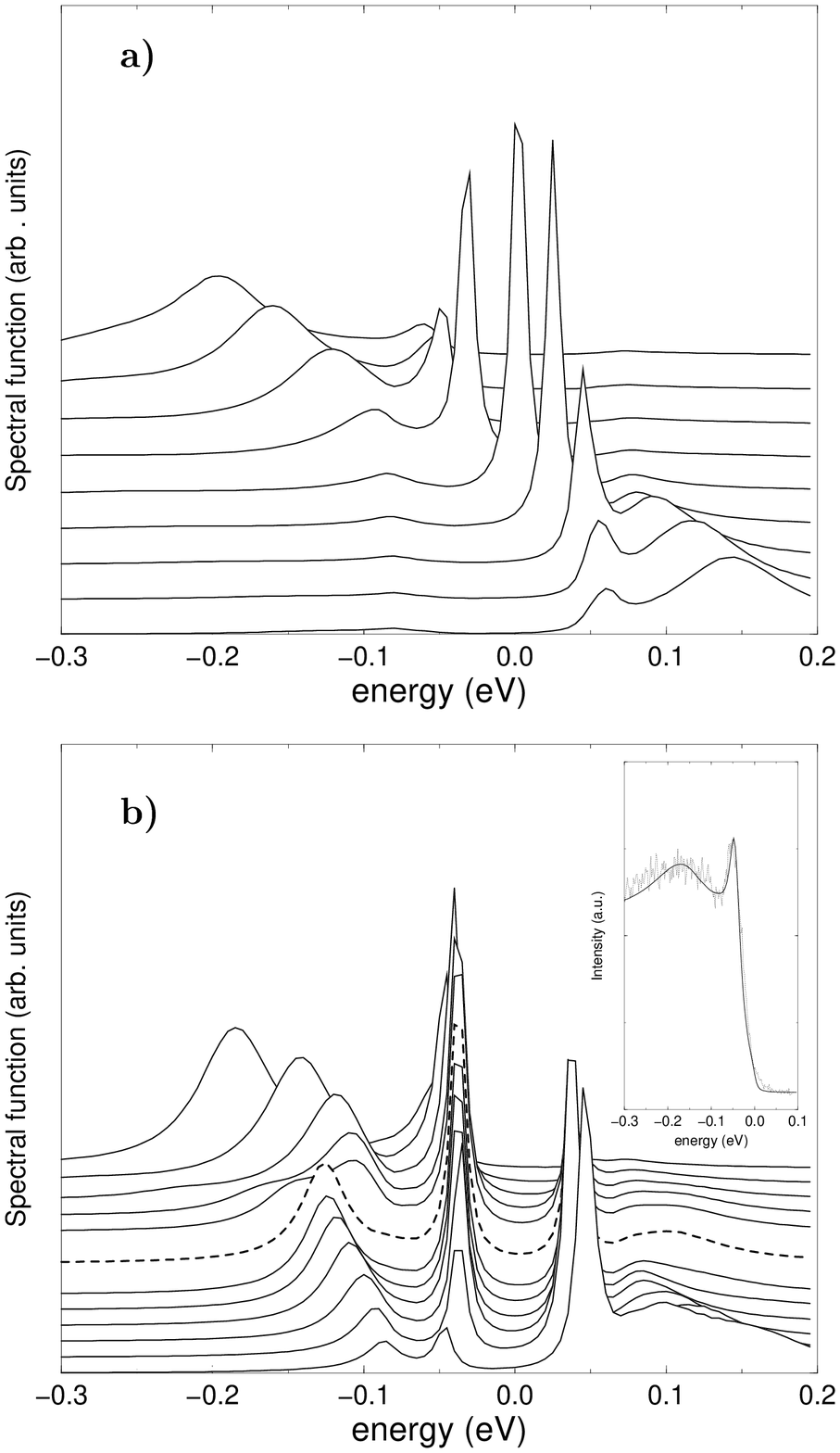,width=7.5cm}}}

{\small FIG. 1. Spectral functions for states from (top to bottom) 
(a) $(0.15\pi,0.36\pi)
\rightarrow (0.5\pi,0.36\pi)$ and (b) $(0.5\pi,0)\rightarrow (\pi,0)
\rightarrow (\pi,0.3\pi)$. The dashed line correpsonds to the M-point.
Parameters: $\omega_0=40meV, \lambda=10meV, \Gamma=0.1, |q_c|=\pi/4$.
Temperature $T=40K$.
Inset:  Spectral function at $(\pi,0)$  
in comparison with experimental data points for an underdoped
sample with $T_c=70K$ measured at 40K.
A step-like function has been added in order to incorporate the
background.
Parameters: $\omega_0=70meV, \lambda=55meV, \Gamma=0.5, |q_c|=\pi/3$.}
\end{figure}

In Figs. 1 (a) and (b) we show the spectral function
for two ``vertical'' cuts, one through the order parameter node at the Fermi
energy $(0.2\pi,0.36\pi)\to (0.5\pi,0.36\pi)$ and the other 
for a cut from $(0.5\pi,0)\to (\pi,0)
\to (\pi,0.3\pi)$. Parameters have been choosen in approximate
analogy to the ones used in \cite{ESCHRIG}. 

The effect of the weight function $J({\bf q})$ is to strongly
reduce the scattering along the diagonals since the states
$({\bf k},\omega)$ and $({\bf k+q_c},\omega-\omega_0)$ 
which are coupled via the CDW fluctuations are very different in energy.
As a consequence the quasiparticles are poorly scattered by the
charge collective modes. Therefore 
the hump feature is suppressed near the node 
and one observes a single dispersive quasiparticle peak at the Fermi energy.
Scanning away from the node towards $(0.2\pi,0.36\pi)$ (Fig. 1a)
the low energy peak rapidly looses weight in favor of the 
hump feature at higher binding energy. This latter broad peak 
arises because moving away from the Fermi surface
the quasiparticles have enough energy to excite charge fluctuations
with ${\bf q} \sim {\bf q}_c$ and are strongly damped. 
In this region of the momentum space the lower-energy narrow 
peak progressively acquires the character of the collective mode while
the hump becomes dispersive as a remnant of the
underlying quasiparticle band. The behavior of the spectral density
along this first cut is rather similar to the case of electrons
coupled to a dispersionless phonon \cite{ENGELSBERG}. In this latter
case, however, the phonons are coupled to the quasiparticles
for any momentum transfer leading to broad incoherent features at high energy
even for quasiparticles on the Fermi surface. On the contrary,
in the present case, the scattering between the quasiparticles near the
node is severely suppressed by the factor $J({\bf q})$ and the incoherent 
hump is nearly absent when the quasiparticle is close to the
Fermi surface (i.e. near the node at $ (0.36\pi,0.36\pi)$).
The switching of the underlying quasiparticle from the 
low-energy peak to the hump structure leads to a break
in the dispersion.  This break of the dispersion is also
observed experimentally \cite{KAMINSKI} however, 
only for momentum scans far from the node. We want to note that 
the dicrepancy of the break-location is a consequence of our    
approximation of a factorized susceptiblity and should be less
pronounced in a more realistic model. An additional possible difficulty
for the present scheme is that the quasiparticles near the node
are only weakly scattered and therefore, according to the
standard Landau theory of Fermi liquids,  they shoud have a damping
rate proportional to $T^2$ and to (binding energy)$^2$. This
Fermi-liquid behavior is violated in recent experiments carried above and
below $T_c$ \cite{VALLA}. However, these experiments were not
carried out at very low temperatures, where a fair comparison can be done
with our $T=0$ analysis, and where heat-transport experiments report
a rather standard Fermi-liquid behavior of the quasiparticles
\cite{TRANSPORTFL} (thereby supporting our finding of well-defined
quasiparticles near the nodes). This issue obviously deserves
further experimental and theoretical investigation. 

Fig. 1b displays the spectral functions for states from 
$(0.5\pi,0) \rightarrow (\pi,0) \rightarrow (\pi,0.3\pi)$. 
The appearance of states with
positive energy is due to coherence effects related to
the finite superconducting gap in this region of momentum space. 
Since the M-point defines the 'hot' region for incommensurate CDW
scattering the superconducting states couple strongly to the CDW mode
and as a result the spectrum (dashed line) is characterized by the 
peak-dip-hump
feature as observed in ARPES. For a suitable choice of parameters we
can obtain a quite satisfactory fit of our curve with experimental data from
photoemission experiments (see inset of Fig. 1b).  

Note that the structure of the hump feature is mainly determined by
the charge-charge correlation function $J({\bf q})$. 
From our fit to the experimental data we can therefore
deduce that in Bi2212 the underlying CDW fluctuations are rather 
short-ranged i.e. the stripe correlations extend over 2-3 unit cells only.
The spectra for the scan $(0.5\pi,0)\rightarrow (\pi,0)$ show an 
interesting feature which is also observed in ARPES (see e.g. 
\cite{CAMPU}, Fig. 3): The sharp peak at low energies is essentially
non-dispersive whereas the hump exhibits a maximum binding energy 
$\omega_{max}$. For energies large than $\omega_{max}$ the hump dispersion
again corresponds to the underlying (normal state) band structure.
The maximum in the hump dispersion is due to the reduced scattering
efficiency (determined by $J({\bf q})$) when moving away form the M-point. 
Since along $(\pi,0) \rightarrow (0,0)$ the underlying dispersion
$\epsilon_k$ initially is rather flat this leads to a shift of the hump to 
lower energies. With $k_x$ becoming smaller the increasing band dispersion
then starts to determine the dispersion of the hump which accordingly shifts
to higher binding energies.

Scanning from M towards $(\pi,\pi)$ the hump moves to lower energies
and looses weight again in agreement with photoemission experiments
\cite{CAMPU,KAMINSKI}.
We want to point out that our result for the spectral functions in Fig. 1 are
rather similar to analogous results of Ref. \cite{ESCHRIG} where
the coupling of electrons to an AF mode has been considered.
Indeed for both AF and incommensurate CDW scattering the 
states around $(\pi,0)$ are the 'hot points', i.e. are most strongly
affected by the scattering. 

This indicates that in the real systems the spin and the charge
fluctuations may well coexist and cooperate in determining the
spectral properties.  Of course for a quantitative comparison with 
experiments one should include both scattering channels. 
On the contrary, since we aim to explore and underline 
the role of charge fluctuations only, in the inset of Fig. 1(b)
we fitted for an illustrative purpose, the
experimental data with the charge channel only. 

\subsection{La$_{2-x}$Sr$_x$CuO$_4$}
In underdoped La$_{2-x}$Sr$_x$CuO$_4$ the incommensurate
magnetic fluctuations display a four-fold pattern around 
$(\pi,\pi)$. For the Nd-doped lanthanum cuprate it has been argued
\cite{TRANQUADA} that this may
be related to two types of twin domains, each with a single stripe
orientation suitably pinned by the underlying LTT lattice structure.
However, this kind of argument no longer holds
in case of underdoped LSCO where both the (tetragonal) a- and b- 
axes are at about 45$^{\circ}$ with respect to the CuO$_6$ tilt direction.
Moreover it has been demonstrated in Refs. \cite{BIRG1,BIRG2} that
the positions of the elastic magnetic peaks in La$_{1.88}$Sr$_{0.12}$CuO$_4$
and excess-oxygen doped La$_2$CuO$_{4+y}$ are shifted off of the
high-symmetry Cu-O-Cu directions by a tilt angle of $\Theta \approx 3^{\circ}$.
For a one-dimensional stripe model this would correspond to one kink every
$\sim 19 $ Cu sites on the charge domain wall. However, as explicitly
stated in Ref. \cite{BIRG2} a grid pattern of orthogonal stripes oriented
along the two orthorombic directions adequately describes the data as well.

\begin{figure}
\hspace*{0.3cm}{{\psfig{figure=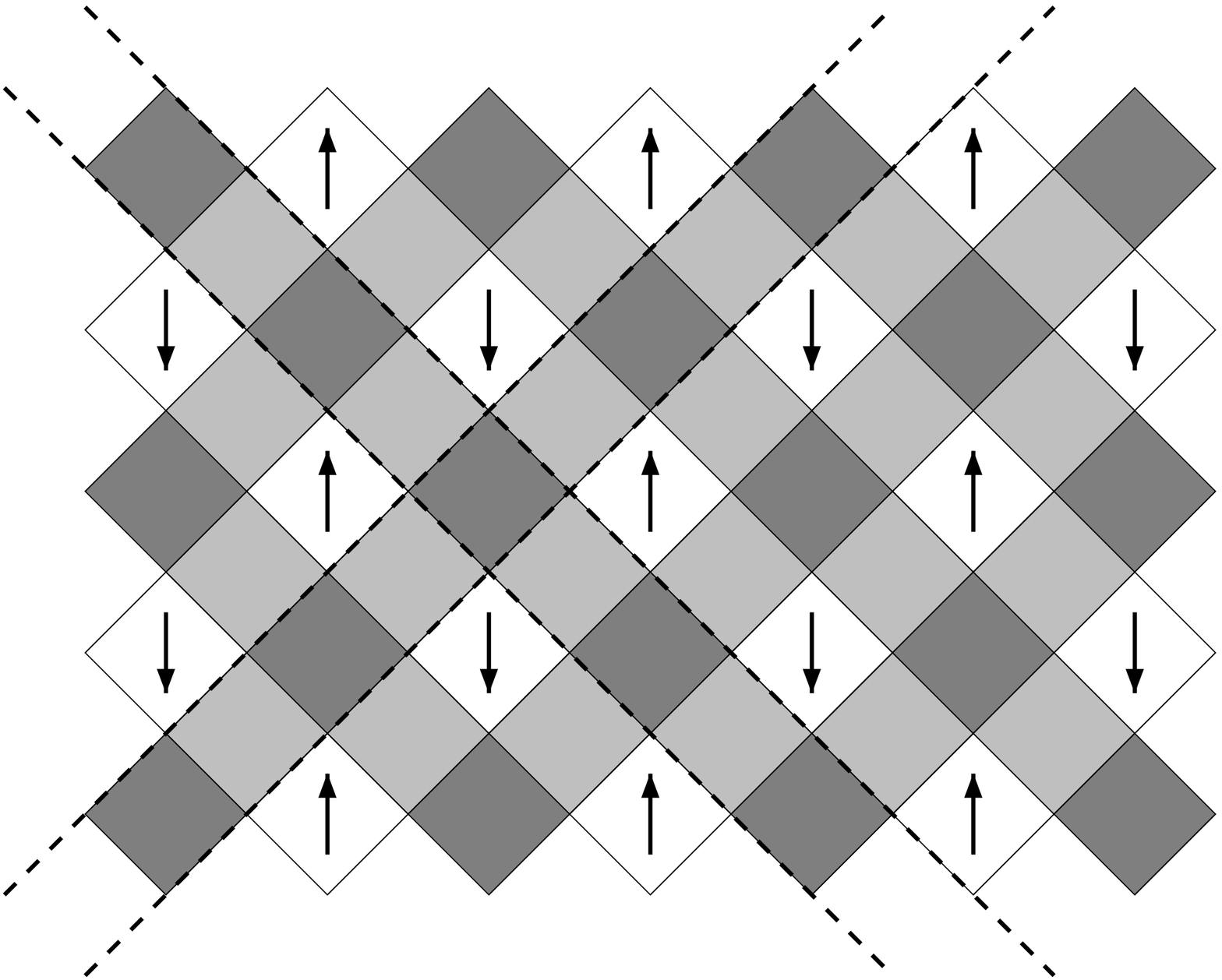,width=7cm,angle=0}}}

{\small FIG. 2. Sketch of the two-dimensional charge- and spin
modulation which is consistent with the AF peak incommensurability of LSCO.
The pattern can be constructed from charge stripes running along the 
diagonal directions as indicated by the dashed lines. The crossing 
of two stripes leads to regions with high charge density (dark squares)
whereas the residual segments of the stripes have intermediate charge
(lightly shaded squares). The white squares illustrate the charge depleted
areas with maximal spin density while the arrows indicate the sign of
the AF order parameter.}
\end{figure}

In the following we therefore examine the consequences of this kind
of two-dimensional dynamic stripe fluctuations on the electronic
structure of LSCO and compare with photoemission experiments.
Denoting the two orthogonal charge scattering vectors 
(which for simplicity we assume to have equal magnitude)
by ${\bf q_c^{\pm}}=q_c(1,\pm 1)$  the charge modulation
can be written as $\rho({\bf R}) \sim \cos(q_c x)\cos(q_c y)$. 
Fig. 2 shows a sketch of the corresponding charge pattern which
can be thought of an array of stripes running along the 
$(1,1)$ and $(1,-1)$ directions respectively. Although our calculations
are restricted on the charge channel we want to note that 
the corresponding pattern of the 
spin order can be constructed by assuming a sign change of the AF 
order parameter upon crossing the stripes perpendicular to their orientation.
This antiphase correlation between the spins across a charge stripe
makes easier the transversal delocalization of the charges in the
stripes and therefore it is energetically favorable. Hence these 
antiphase domain walls which are 
observed in Nd-doped LSCO around 1/8 of filling \cite{TRAN}
seem also to be a rather common feature of the 
theoretically investigated stripe structures. As a consequence,
in our charge-spin structure, the spin order follows the relation 
$\Delta^{spin}({\bf R}) \sim \cos\lbrack q_c(x+y)/2\rbrack
\cos\lbrack q_c(x-y)/2\rbrack \sim \cos({\bf q_s^+ R})+\cos({\bf q_s^- R})$
where the spin scattering vectors are given by ${\bf q_s^{+}}=q_c(1,0)$
and ${\bf q_s^{-}}=q_c(0,1)$ respectively.
Thus this kind of two-dimensional scattering displays a deviation between
charge- and spin direction of 45$^{\circ}$ where the latter is compatible
with the results of neutron scattering experiments \cite{YAM,ARAI}. 
Only in case of a LTT distortion both charge- and spin scattering collapse 
into a single direction leading to a reduction of T$_c$ \cite{TRAN}.

\begin{figure}
\hspace*{0.1cm}{{\psfig{figure=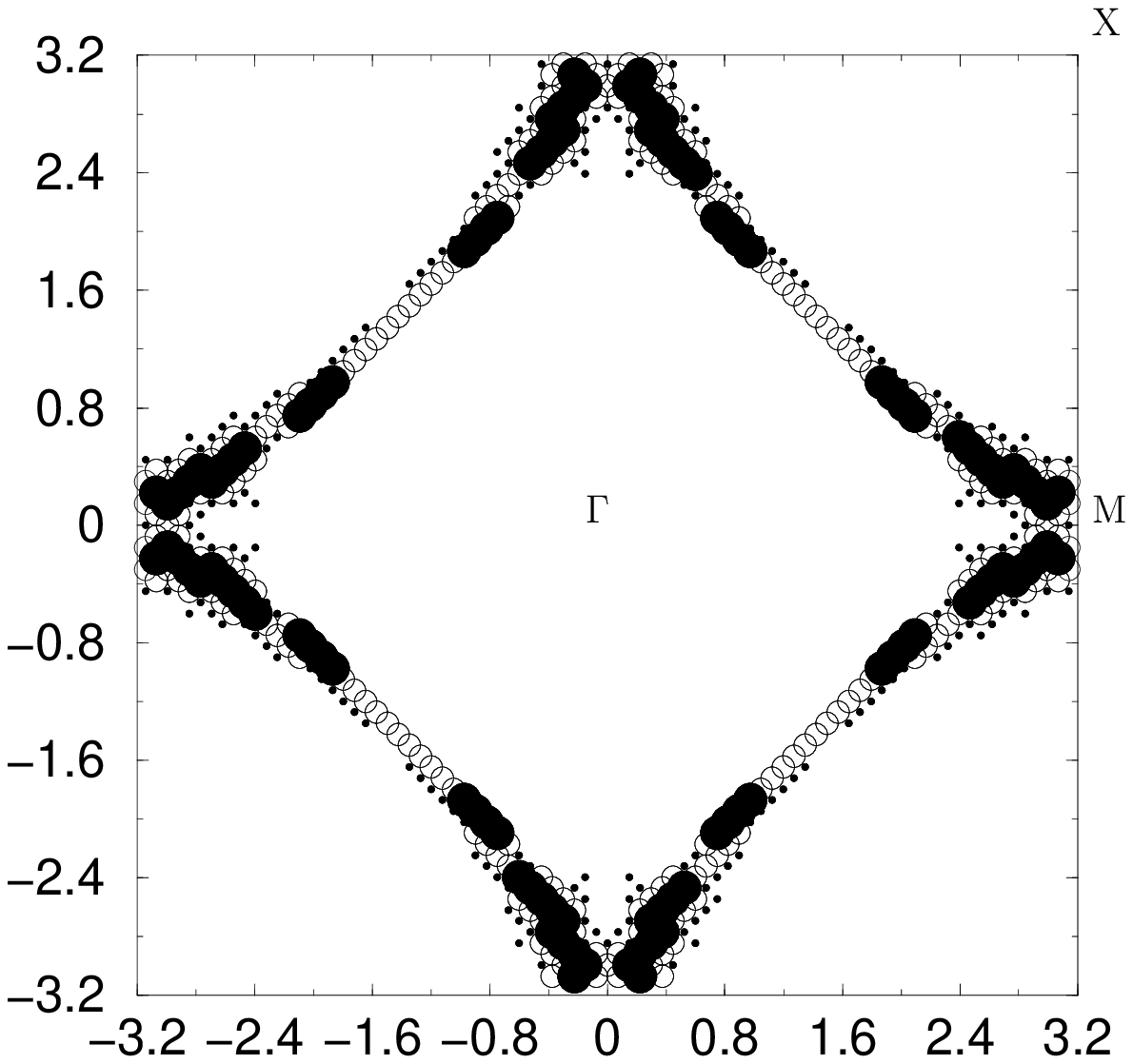,width=7.5cm,angle=0}}}
{\small FIG. 3. Fermi surface for a system with two-dimensional diagonal CDW
modulation ($q_c=\pi/5, \lambda=0.01$).
The plot is for temperature $T=100K$. The spectral weight has been integrated
over an energy window of 30meV around E$_{F}$. Intensities:
$I>50\%$: full points, $10\%<I<50\%$: circles, $1\%<I<10\%$:
small dots.}
\end{figure}

As in the case of Bi2212 we restrict in the following on the charge
channel and as discussed above we consider scattering 
along the ${\bf q_c^{\pm}}=(1,\pm 1)$ directions.
To gain some qualitative insight we consider first the static 
non-superconducting case inspired by the approach described in 
Ref. \cite{GOETZ} where we have derived an effective interaction
$\frac{1}{2N}\sum_q V_{q} \rho_{q}\rho_{-q}$ for charge carriers
close to an incommensurate CDW instability. Upon factorizing the interaction
the two-dimensional eggbox-type charge modulation can be implemented by  
$<\rho_{q}>^{egg}=<\rho_{q}>\lbrack \delta_{q,q_{c}^{+}}
 + \delta_{q,q_{c}^{-}}\rbrack$  which in the static limit 
identifies the coupling parameter $\lambda$ with the order parameter of the 
CDW $V({\bf q_c})<\rho_{\bf q_c}>$. The resulting one particle hamiltonian
can now be diagonalized and as a result we show
in Fig. 3 the Fermi surface for an eggbox-type charge modulation
with  scattering vectors ${\bf q_c}=\pi/5(1,\pm 1)$. 

Obviously diagonal CDW scattering strongly reduces the spectral intensity
of the Fermi surface along the $\Gamma \rightarrow X$ direction. 
This is connected with the fact that along the diagonals 
the scattering vector is parallel to the contours of the bare bandstructure.
As a consequence the CDW scattering is most efficient in these regions 
since ${\bf q_c}$ can connect states with equal energies 
leading to a redistribution of spectral weight near ($\pi/2,\pi/2$).
Note that since the band disperses rather rapidly along $\Gamma \rightarrow X$
the vertical scattering we have adopted in case of Bi2212 
would leave the electronic
structure in this direction nearly unchanged and would not reproduce the 
experimental suppression of spectral weight along the 
$\Gamma \rightarrow X$ direction \cite{INO,INO2,ZHOU}. 

Quite obviously our simplified approach, where the spin degrees of freedom 
are neglected, cannot provide a quantitative description of the 
Fermi surface. However, we notice that a model based on
spin degrees of freedom only,  constrained by the neutron experiments to
have hot spots near the M points could hardly account for the 
suppression of spectral weight along the $\Gamma \to X$ direction.

\begin{figure}
\hspace*{0.5cm}{{\psfig{figure=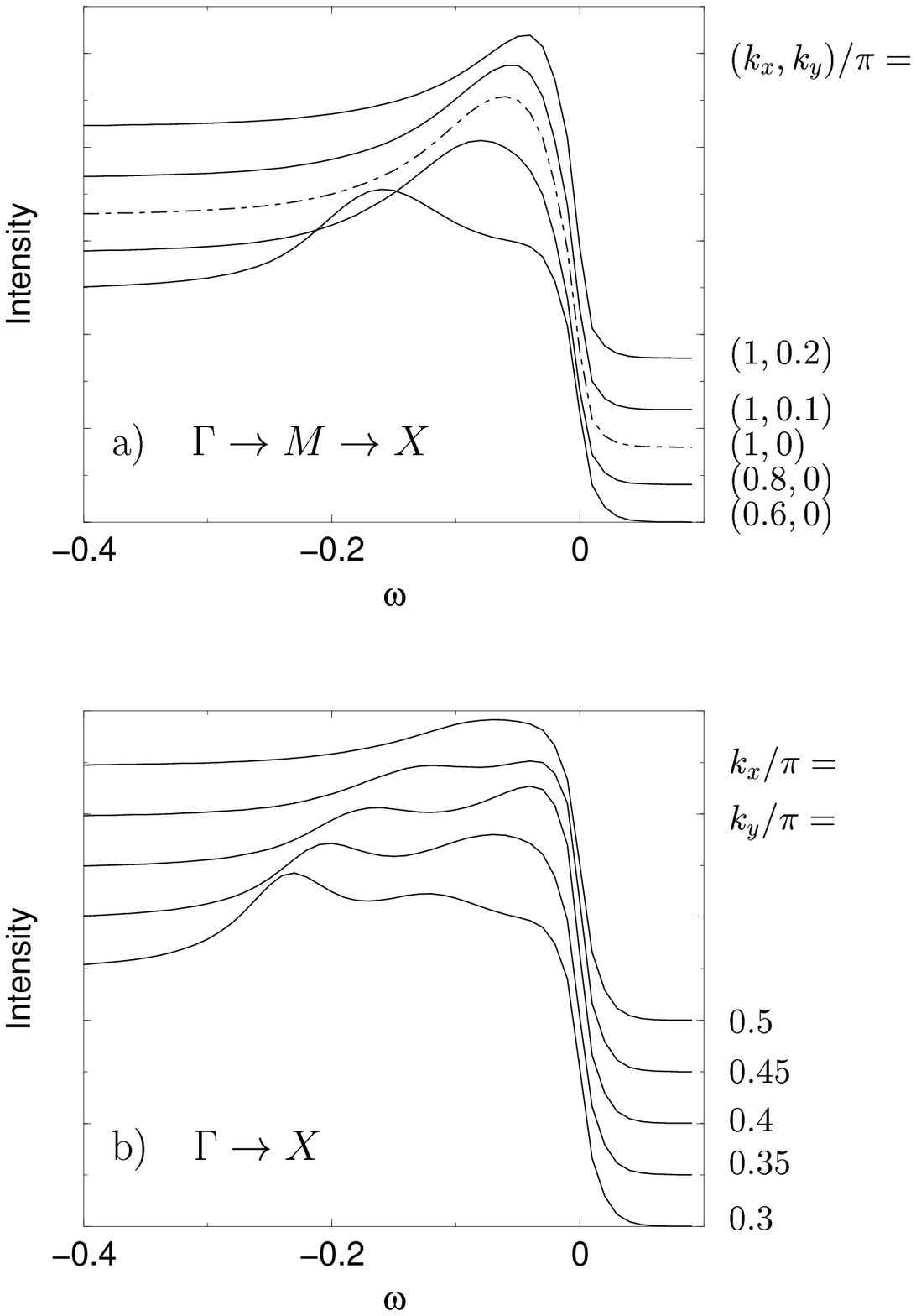,width=7.5cm,angle=0}}}
{\small FIG. 4. Photoemission spectra for diagonal two-dimensional CDW 
scattering. As in the case of Bi2212 we have added a step-like function
in order to model the background contribution.
Parameter: $\omega_0=30$meV, $\Gamma=0.01$, 
${\bf q_c}= \pi/5 (1,\pm1)$.}
\end{figure}

Let us now extend the photoemission lineshape analysis to the 
superconducting state including
dynamic incommensurate CDW scattering. We have found that for the LSCO compound
a linear frequency distribution $g(\nu)=\frac{2\nu}{\omega_0^2}
\Theta(\omega_0-\nu)$ up to a cutoff energy $\omega_0$ 
is more appropriate consistent with the experimental observation
of low energy magnetic fluctuations in this compound.
Fig. 4 shows selected energy distribution curves along 
$\Gamma \rightarrow M \rightarrow X$ (Fig. 4a) and $\Gamma \rightarrow X$
(Fig 4b).

In the $(0,0) \rightarrow (\pi,\pi)$ 
direction (Fig. 4b) one observes only a broad and weak incoherent
feature which disappears beyond $k_x=k_y \sim 0.45 \pi$ corresponding
to the FS crossing of the underlying bare bandstructure.
In contrast the spectra around $(\pi,0)$ (Fig. 4a) are composed of
an also broadened but still well pronounced quasiparticle peak.
Due to the small cutoff energy $\omega_0$ in comparison with Bi2212
and the additional linear frequency distribution the hump feature
is no longer separated from the sharp peak but both collapse into
the broadened peak shown in Fig. 4b.
Note that we were forced to use 
a rather large charge-charge correlation length 
($\sim 100$ lattice constants) in order to effectively suppress the
quasiparticle peak along $(0,0) \rightarrow (\pi,\pi)$. This 
finding needs to be commented. On the one hand
it is surely conceivable that the poor quality of the surface
in 214 materials is responsible for the width of the 
photoemission peaks. In this case this would be a spurious effect
as far as the bulk electronic properties are concerned and our
need of a large charge-charge correlation length would be fictitious.
On the other hand the authors of Refs. \cite{INO,INO2,ZHOU} claim
that a careful attention was paid to the surface treatment.
Moreover, recent photoemission experiments \cite{YOSHIDA} on
overdoped LSCO have revealed a sharp spectral feature along the
diagonal direction that is comparable to that of Bi2212. From this
one would not expect strong disorder effects in the underdoped samples which
should improve with decreasing Sr content. 
In this regard our finding of a large
charge-charge correlation length could be taken seriously at least from
a qualitative point of view. The presence of rather long-ranged
charge-ordering correlations would rather naturally
correspond to the much more visible presence of stripe correlations
in 214 materials. Indeed, but for some inelastic neutron scattering
\cite{MOOK} and some NMR/NQR \cite{BRINKMAN,MEHRING} experiments in YBCO,
the evidence for stripes is rather clear in 214 materials only, while it 
remains more elusive in the other classes of cuprates.

\section{Conclusion}
We have demonstrated that the coupling of a superconducting system
to dynamical incommensurate CDW fluctuations can account for the
observed ARPES photoemission lineshapes in both Bi2212 and LSCO
materials. According to our analysis the vertically oriented 
CDW fluctuations in Bi2212 are rather fast and short ranged 
leading to the experimentally observed dip-hump structure around the M-points 
whereas the lineshape along the diagonals is hardly affected by the 
scattering. 

We emphasize again here that our analysis shares many similarities
(both in the technical framework and in the results) with the
one carried out in Ref. \cite{ESCHRIG} 
 (the renormalization function $Z(\omega)$ obtained from our approach 
is quasi identical to Fig. 1 in Ref. \cite{ESCHRIG}). 
Therefore our work is not in contrast with Ref. \cite{ESCHRIG}
but rather complementar since the stripe fluctuations we consider here
are anharmonic CDW's with the strong correlation playing a relevant role
in the hole-poor regions, where antiferromagnetism is quite pronounced.
For the sake of simplicity we only investigated the charge sector
of the fluctuations. Nevertheless, the effects of spin \cite{ESCHRIG} 
and charge (this work) fluctuations are similar indicating 
that these fluctuations may easily coexist and
that there is no competition but rather cooperation between the
two sectors. Of course a long way is still to be followed in order to
integrate the two approaches by formalizing the interplay between
charge and spin degrees of freedom. In particular an open relevant issue
is to establish how the critical properties related to the charge instability
are mirrored in the spin criticality. This intriguing but difficult
issue is definitely beyond the present scope.

In case of LSCO the photoemission data can be described by  
a two-dimensional, diagonally oriented CDW which is slowly fluctuating
and characterized by a rather large correlation length. This is consistent
with the idea of long-range AF order coexisting with 
superconductivity in the 
214 systems and also with the more pronounced tendency to form static
charge textures in these materials. 
We have shown that within this scenario the QP along
the $(0,0) \rightarrow (\pi,\pi)$ direction can be effectively suppressed
whereas around the M-points a broad quasiparticle peak persists.
Finally we want to emphasize that since $\Delta_{SC}$ enters the calculation
as an input parameter, our model does not include the phase fluctuations
of the particle-particle pairs.  These fluctuations should play a
major role in destroying the quasiparticle peak above T$_c$ \cite{FRANZ}.
However, the measurements in Refs. \cite{INO,INO2,ZHOU} on LSCO were done
at rather low temperatures (T=15K) where such phase fluctuations are
uneffective justifying our simplified model for the photoemission spectra.

At the present stage of the research in the field of superconducting
cuprates our  simplified analysis is particularly urgent since it 
illustrates that charge fluctuations can generically be 
quite effective sources of scattering between the quasiparticles
providing spectral features which are not in contrast with experiments
both in LSCO systems, where there seem to be little doubt that
(dynamical) stripes are present, and in Bi2212 compounds, where
no stringent evidence for stripe fluctuations is available. In this
regard we show that different kinds of charge fluctuations 
can operate in different materials. 
Moreover, in the present situation, where even the
experimental situation is rapidly evolving, we believe that 
our results can be useful in keeping the community flexible
enough to effectively determine the intricate
physical mechanisms of the superconducting cuprates.

\acknowledgments 
We would like to thank J. Mesot for providing the experimental
data in Fig. 1. We also greatfully acknowledge useful discussions 
with S. Caprara, C. Castellani and C. Di Castro.

\end{multicols}

\end{document}